Superconducting phase diagram of Na$_x$CoO$_2$·$y$H$_2$O

Hiroya Sakurai[a,b,*], Kazunori Takada[c], Takayoshi Sasaki[c], Eiji Takayama-Muromachi[a]

[a] Superconducting Materials Center, National Institute for Materials Science, Namiki 1-1, Tsukuba, Ibaraki 305-0044, Japan

[b] International Center for Young Scientists, National Institute for Materials Science, Namiki 1-1, Tsukuba, Ibaraki 305-0044, Japan

[c] Advanced Materials Laboratory, National Institute for Materials Science, Namiki 1-1, Tsukuba, Ibaraki 305-0044, Japan

Abstract

We synthesized Na$_x$(H$_3$O)$_z$CoO$_2$·$y$'H$_2$O samples with various Na/H$_3$O ratios but with the constant Co valence of $s$ = +3.40, and measured their magnetic properties to draw phase diagrams of the system. The superconductivity is very sensitive to the Na/H$_3$O ratio. With varying $x$ under fixed $s$ of +3.40, magnetically ordered phase appears in the intermediate range of $x$ sandwiched by two separated superconducting phases, suggesting that the superconductivity is induced by moderately strong magnetic interactions. In the vicinity of the magnetic phase, transition from the superconducting state to the magnetically ordered state was induced by applying high magnetic field. This transition is of the second order, at least, above 1.8 K. The upper-critical field is expected to be much higher than the Pauli limit for a phase located far away from the magnetic phase regarding the Na/H$_3$O ratio.

PACS codes: 74.62.-c, 74.62.Bf

Keywords: $Na_xCoO_2 \cdot yH_2O$; magnetic order; phase diagram

Sodium cobalt oxyhydrate superconductor, $Na_xCoO_2 \cdot yH_2O$, has been intensively studied since its discovery [1]. However, its phase diagram is still controversial with conflicting $x$-$T$ diagrams ($T$: temperature) reported thus far [2, 3], suggesting that there is a hidden compositional parameter in the system. Indeed, it has been found that a significant amount of oxonium ions exist in the compound and the chemical formula should be expressed as $Na_x(H_3O)_zCoO_2 \cdot y'H_2O$ rather than $Na_xCoO_2 \cdot yH_2O$ [4]. The presence of the oxonium ions is crucial because the Co valence, $s$, cannot be estimated only from the Na content any more. Considering the oxonium ions, a revised phase diagram has been recently proposed with a dome shape of superconducting region of ~+3.2 < $s$ < +3.4 (the optimal $T_c$ of 4.8 K at +3.24 < $s$ < +3.35) [5]. However, this diagram seems unlikely because the cobalt valence is much lower than $s$ = +3.42 to +3.48 reported previously [4, 6-8]. On the other hand, there is no consensus for the $H$-$T$ phase diagram ($H$: magnetic field), as well. With magnetic field along the $ab$-plane, the upper-critical field ($H_{c2}$) has been estimated, but the values reported thus far are scattered very largely: ~8 T [9, 10] to over 15 T [11-14]. Since the Pauli limit of $1.84 \cdot T_c$ ~ 8 T [15] is one of criteria to deduce the spin state of the Cooper pairs, it is very important whether $H_{c2}$ is around 8 T or significantly larger than it.

The samples were synthesized by soft-chemical method from the precursor of $Na_{0.7}CoO_2$ with a single batch, which was made from $Na_2CO_3$ and $Co_3O_4$ as reported elsewhere [16]. A certain amount of the precursor (1 g) was immersed in 5 vol.% $Br_2/CH_3CN$ (35 ml) to deintercalate the $Na^+$ ions, and then was immersed in distilled water (400ml) to intercalate water molecules. Before filtration, 0.1 M HCl aqueous solution ($v$ ml) was added to the solution to control the chemical composition of the sample. The samples reported here are the same as those in the previous report [17]. As seen in Fig. 1, x-ray diffraction patterns indicate that the sample is obtained as single phase for $v \leq 10$ ml, but secondary phase of CoOOH or its related compound appears for $v \geq 12$ ml. After the samples were stored in the air with 70% humidity for more than 3 weeks, chemical analyses by inductive-coupled plasma atomic emission spectroscopy and redox titration were performed for each sample almost simultaneously with magnetic measurements. The results of the chemical analysis are shown in Table 1 as well as the $c$-axis lengths and $T_c$'s estimated from magnetic susceptibility data under 0.001 T. The $Na^+$-ion content decreases obviously with increasing $v$ whereas the Co valence is almost constant to be about +3.40. This result is reasonable considering the isovalent ion exchange of $H_3O^+$ for $Na^+$ which is expected to be promoted by the addition of the HCl solution [17]. The increasing $c$-axis supports this exchange reaction because the $Na^+$ ion is smaller than the $H_3O^+$ ion.

It is notable that the superconductivity disappears for the intermediate range of $v$, i.e., for $v$ = 2, 4, and 6 ml in spite of monotonous changes of the Na content and the $c$-axis length throughout the $v$ range of $0 \leq v \leq 10$ ml. The magnetic susceptibilities of these non-superconducting samples show an anomaly at about 6 K as seen in Fig. 2(a). This anomaly obviously corresponds to a magnetic transition that has been observed by $^{59}$Co nuclear quadrupole resonance for a non-superconducting sample [18]. Taking into account the non-superconducting magnetic phase, the phase diagram may be depicted as Fig. 3(a). The superconducting region is separated into two parts of *SC*1 and *SC*2 sandwiching the magnetically ordered *M* phase, suggesting that the superconductivity is induced by moderately strong magnetic interactions.

It is important to notice for Fig. 3(a) that the Co valence is kept constant to be +3.40 in this section of phase diagram. Thus, the magnetic interaction is affected critically by the ion exchange of $H_3O^+$ for $Na^+$ even when the Co valence is constant. At the present stage, it is most likely that the superconductivity and the magnetic ordering are dominantly undertaken by holes or electrons on pocket-like Fermi surface near K-point [19] or at Γ-point [20], respectively. The size of the pocket seems to be very susceptible to the thickness of the $CoO_2$ layer [21, 22]. By the ion exchange of $Na^+$ for $H_3O^+$, the layer is expected to become thinner because oxygen ions of the layer are less attracted by the $Na^+/H_3O^+$ ions with increasing distance between them (increasing the *c*-axis length). The importance of the pocket-like Fermi surface has been pointed out by theoreticians [20, 23], and its existence is consistent with Co-valence dependence of $T_c$ [24].

The present study indicates that two compositional parameters should be specified to draw a phase diagram among the Na content *x*, $H_3O$ content *z* and Co valence *s* ($s = 4 − x − z$). Only one parameter was taken into account in the previous phase diagrams [2, 3, 5], and thus they reflect some cross-sections of three-dimensional *x*-*s*-*T* phase diagram. The *SC*1 and *SC*2 phases are possibly connected with each other somewhere in the whole *x*-*s*-*T* phase diagram.

The magnetic susceptibility measured under 7 T is shown in Fig. 2(b), which is important to solve the inconsistency in the *H*-*T* phase diagram. The *v* = 0 ml sample whose composition (*x* = 0.35) is located very close to the *M* phase in Fig. 2(a), shows a magnetic ordering under 7 T instead of the superconducting transition, similar to the samples with *v* = 2, 4, or 6 ml. Namely, a transformation from a superconducting phase to a magnetic phase is induced in the $v = 0$ ml sample by magnetic field. Such a transformation is rare, but it seems natural to occur in the present case because magnetic field should destabilize the superconducting state while can stabilize the magnetically ordered state. In other words, it is not curious that the $M$ phase extends to a wider $x$ region under a high magnetic field.

In order to determine the phase boundary between the $SC$1 and the $M$ phases, the magnetic susceptibility was measured under various fields for the $v = 0$ ml sample, as seen in Fig. 4(a). Two anomalies are clearly recognized for each curve: for example, around 2.8 K and 5.7 K under 5.5 T. Thus, the $H$-$T$ phase diagram can be depicted as in Fig. 3(b). The transformation can be also detected in isothermal magnetization curves at 1.8 K and 3 K as seen in Fig. 4(b), although the transformation field is a little smaller than that expected from Fig. 4(a). These behaviors have intrinsic origin, not extrinsic one such as sample inhomogeneity, because the steep decrease in $T_c$ above 5 T indicates that the magnetic ordering suppresses the superconducting transition, and because the phenomenon does not look like a crossover but a transition as suggested by the $M$-$H$ curves. No anomaly was observed in the $M$-$H$ curve of the $v = 6$ ml sample up to 7 T. The transition is of the second order, at least, above 1.8 K since no significant hysteresis is seen.

At last, we give a comment on the $H$-$T$ phase diagrams reported thus far. The smaller $H_{c2}$ of about 8 T seems to be due to the transition from the superconducting state to the magnetic state. Indeed, the $SC$1-$M$ boundary in the $H$-$T$ diagram in Fig. 3(b) seems to agree well with the field dependence of $T_c$ by which the smaller $H_{c2}$ was estimated [9, 10]. When a composition of a phase is far away from the $SC$1(or $SC$2)-$M$ boundary, the higher $H_{c2}$ is expected without appearance of the $M$ phase. Actually, the susceptibility of the $v$ = 10 ml sample under 7 T shows downturn at 3.5 K as a clear sign of superconductivity (Fig. 2(b)) which is significantly higher than ~2.6 K or 1.7 K observed in the samples with $H_{c2}(0)$ ~ 8 T [9, 10]. $T_c$ of the $v$ = 0 ml sample may be even higher than 3.5 K because it is difficult to estimate exact field dependence of $T_c$. Decrease of the susceptibility due to the superconducting diamagnetism seems to start at temperature higher than 3.5 K in Fig. 2(a), though it is hidden by the increasing back-ground susceptibility and hard to be specified.

In summary, we synthesized $Na_x(H_3O)_zCoO_2 \cdot y'H_2O$ samples with various Na/H$_3$O ratios but with the constant Co valence of $s$ = +3.40, and measured their magnetic properties to draw phase diagrams of the system. The superconductivity is very sensitive to the isovalent ion exchange of $H_3O^+$ for $Na^+$, indicating that not one but two compositional parameters are needed to be specified. With varying $x$ under fixed $s$ of +3.40, magnetically ordered phase appears in the intermediate range of $x$ sandwiched by two separated superconducting phases. This unique phase diagram strongly suggests that moderately strong magnetic interaction is crucial to the superconductivity. The magnetic interaction is likely undertaken by carriers on the pocket-like Fermi surface. It was also proved for a superconducting phase located close to the magnetic phase regarding the composition that it is transformed to the magnetic phase by applying high magnetic field. The transition is of the second order at least above 1.8 K. $H_{c2}$ is expected to be significantly higher than the Pauli limit for a superconducting phase located far away form the magnetic phase.

**Acknowledgements**

We would like to thank S. Takenouchi and K. Kosuda for the chemical analyses.

This work is partially supported by Grants-in-Aid from JSPS and MEXT (16340111, 16076209) and by CREST, JST.

**References**

[1] K. Takada, H. Sakurai, E. Takayama-Muromachi, F. Izumi, R. A. Dilanian, T. Sasaki, Nature (London) 422 (2003) 53.

[2] R. E. Schaak, T. Klimczuk, M. L. Foo, R. J. Cava, Nature (London) 424 (2003) 527.

[3] D. P. Chen, H. C. Chen, A. Maljuk, A. Kulakov, H. Zhang, P. Lemmens, C. T. Lin, Phys. Rev. B 70 (2004) 024506.

[4] K. Takada, K. Fukuda, M. Osada, I. Nakai, F. Izumi, R. A. Dilanian, K. Kato, M. Takata, H. Sakurai, E. Takayama-Muromachi, T. Sasaki, J. Mater. Chem. 14 (2004) 1448.

[5] C. J. Milne, D. N. Argyriou, A. Chemseddine, N. Aliouane, J. Veira, S. Landsgesell, D. Alber, Phys. Rev. Lett. 93 (2004) 247007.

[6] M. Karppinen, I. Asako, T. Motohashi, H. Yamauchi, Chem. Mater. 16 (2004) 1693.

[7] M. Bañobre-López, F. Rivadulla, R. Caudillo, M. A. López-Quintela, J. Rivas, J. B. Goodenough, Chem. Mater. 17 (2005) 1965.

[8] K. Takada, M. Osada, F. Izumi, H. Sakurai, E. Takayama-Muromachi, T. Sasaki, Chem. Mater. 17 (2005) 2034.

[9] F. C. Chou, J. H. Cho, P. A. Lee, E. T. Abel, K. Matan, Y. S. Lee, Phys. Rev. Lett. 92 (2004) 157004.

[10] T. Sasaki, P. Badica, N. Yoneyama, K. Yamada, K. Togano, N. Kobayashi, J. Phys. Soc. Jpn. 73 (2004) 1131.

[11] P. Badica, T. Kondo, K. Togano, K. Yamada, cond-mat/0402235.

[12] R. Jin, B. C. Sales, S. Li, D. Mandrus, cond-mat/0410517.

[13] M. M. Maśka, M. Mierzejewski, B. Andrzejewski, M. L. Foo, R. J. Cava, T. Klimczuk, Phys. Rev. B 70 (2004) 144516.

[14] H. Sakurai, K. Takada, S. Yoshii, T. Sasaki, K. Kindo, E. Takayama-Muromachi, Phys. Rev. B 68 (2003) 132507.

[15] A. M. Clogston, Phys. Rev. Lett. 9 (1962) 266.

[16] H. Sakurai, S. Takenouchi, N. Tsujii, E. Takayama-Muromachi, J. Phys. Soc. Jpn. 73 (2004) 2081.

[17] H. Sakurai, K. Takada, T. Sasaki, E. Takayama-Muromachi, J. Phys. Soc. Jpn. 74 (2005) 2909.

[18] Y. Ihara, K. Ishida, C. Michioka, M. Kato, K. Yoshimura, K. Takada, T. Sasaki, H. Sakurai, E. Takayama-Muromachi, J. Phys. Soc. Jpn. 74 (2005) 867.

[19] D. J. Singh, Phys. Rev. B 61 (2000) 13397.

[20] K. Kuroki, S. Onari, Y. Tanaka, R. Arita, T. Nojima, cond-mat/0508482.

[21] M. Mochizuki, Y. Yanase, M. Ogata, J. Phys. Soc. Jpn. 74 (2005) 1670.

[22] K. Yada, H. Kontani, J. Phys. Soc. Jpn. 74 (2005) 2161.

[23] Y. Yanase, M. Mochizuki, M. Ogata, J. Phys. Soc. Jpn. 74 (2005) 430, and references therein.

[24] H. Sakurai, N. Tsujii, O. Suzuki, H. Kitazawa, G. Kido, K. Takada, T. Sasaki, E. Takayama-Muromachi, unpublished.

Table 1  Na content, Co valence, $c$-axis length, and $T_c$ of the samples.

| $v$ (ml) | Na content | Co valence | $c$ (Å) | $T_c$ (K) |
|---|---|---|---|---|
| 0 | 0.350 | 3.41 | 19.717 | 4.3 |
| 2 | 0.346 | 3.40 | 19.724 | < 1.8 |
| 4 | 0.346 | 3.41 | 19.752 | < 1.8 |
| 6 | 0.336 | 3.40 | 19.775 | < 1.8 |
| 8 | 0.322 | 3.40 | 19.820 | 4.1 |
| 10 | 0.302 | 3.36 | 19.810 | 4.3 |

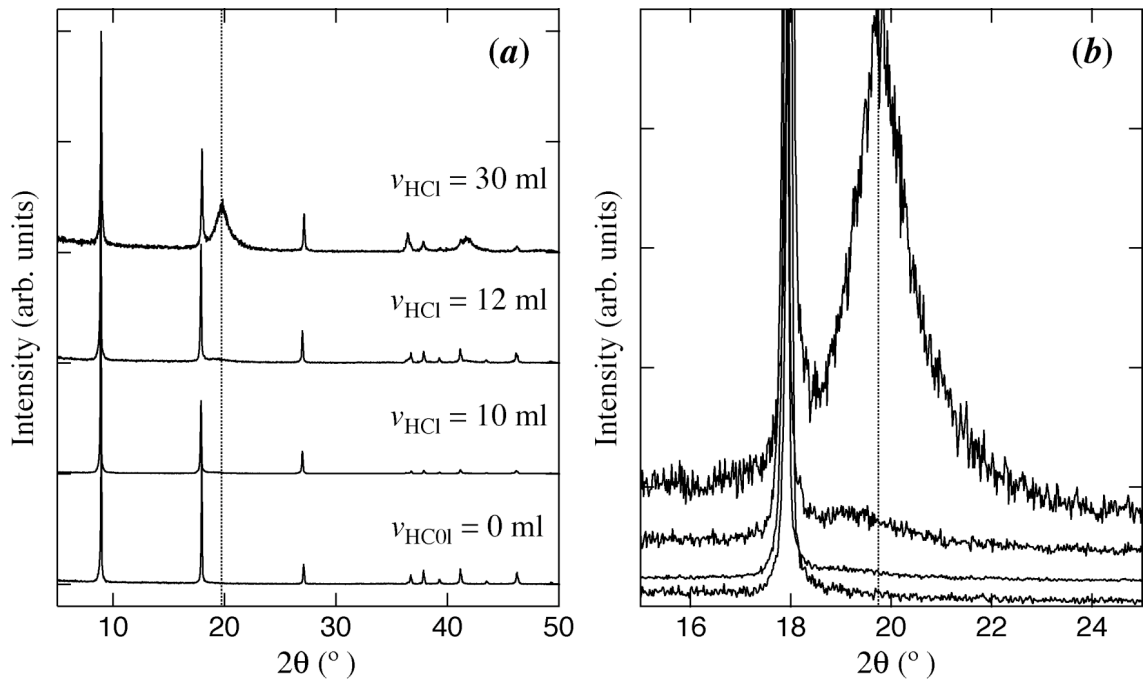

Fig. 1  X-ray diffraction patterns of the $v$ = 0, 10, 12, and 30 ml samples. The right panel is the enlarged copy of the left one. The broken lines represent a basal reflection of CoOOH.

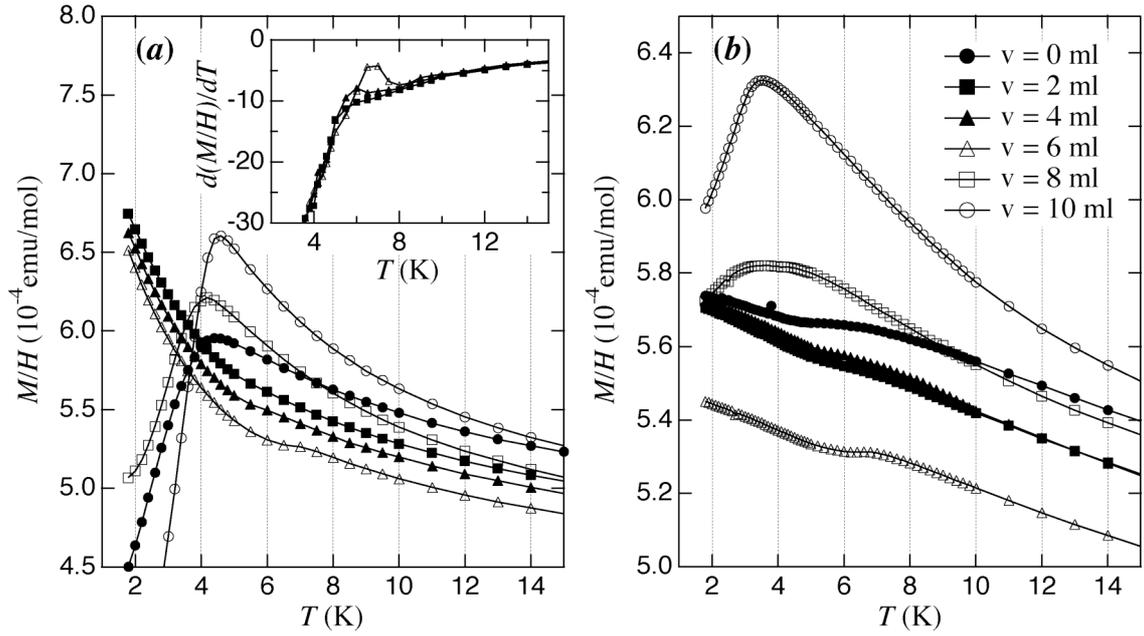

Fig. 2 Magnetic susceptibility measured under 1 T (a) and 7 (T). The inset of the upper panel shows the differentiated susceptibilities of the $v$ = 2, 4, and 6 samples.

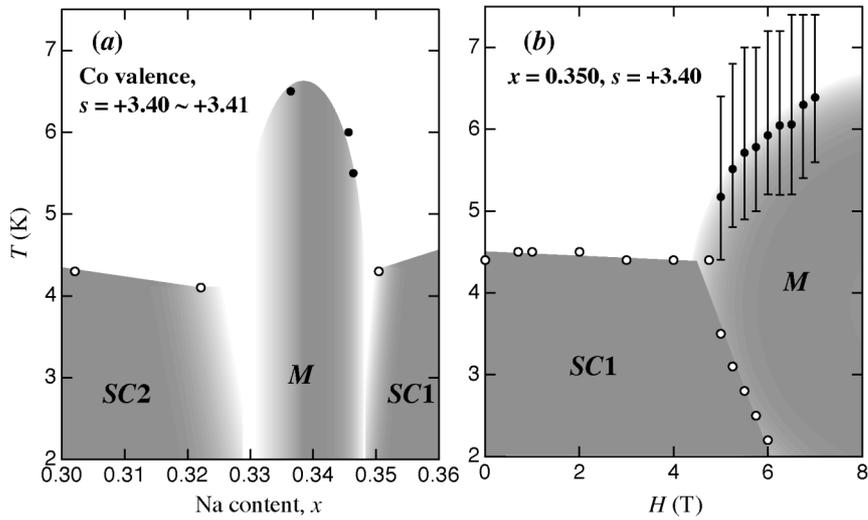

Fig. 3 The $x$-$T$ phase diagram (a) and $H$-$T$ phase diagram (b). The open and closed circles represent $T_c$ and the magnetic ordering temperature, respectively.

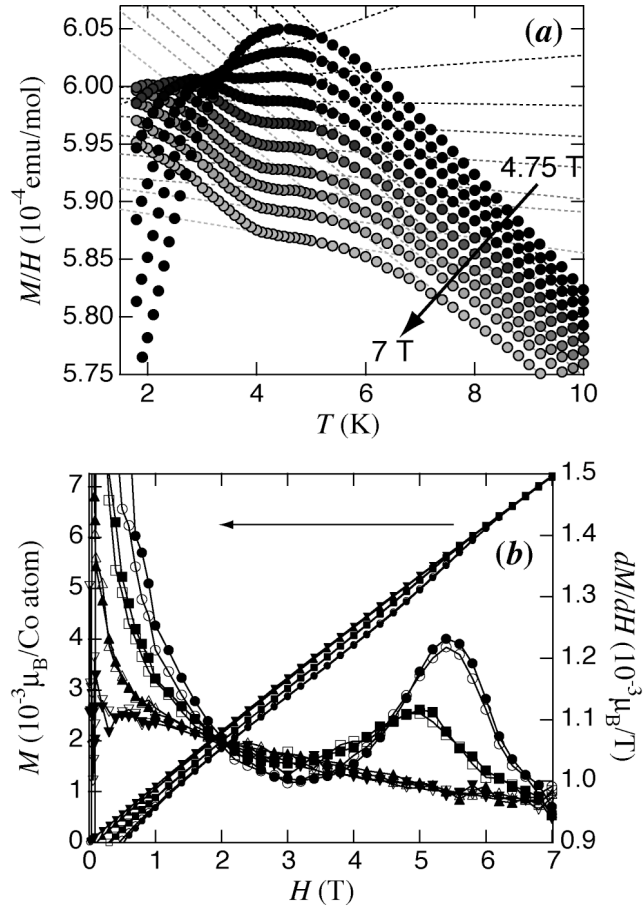

Fig. 4 (a) Magnetic susceptibility under the fields between 4.75 T and 7 T with each interval of 0.25 T and (b) magnetization curves at 1.8 K (circles), 3 K (squares), 4 K (triangles), and 5 K (inverse triangles) of the $v = 0$ ml sample. In the lower panel, the closed and open markers represent the data collected under increasing and decreasing fields, respectively.